\documentclass[%
 reprint,
 amsmath,amssymb,
 aps,
]{revtex4-2}

\usepackage{graphicx}
\usepackage{dcolumn}
\usepackage{bm}
\usepackage{hyperref}
\usepackage{xcolor}

\begin{document}

\preprint{none}

\title{Is the muon a third family lepton?}

\author{Giacomo Cacciapaglia}
\email{g.cacciapaglia@ip2i.in2p3.fr}
\affiliation{
Institut de Physique des 2 Infinis de Lyon (IP2I), IN2P3/CNRS, UMR5822, rue Enrico Fermi, Villeurbanne, France}
\author{Aldo Deandrea}%
\email{deandrea@ip2i.in2p3.fr}
\affiliation{
Universit\'e de Lyon, F-69622 Lyon, France; Universit\'e Lyon 1,
Villeurbanne, IN2P3/CNRS, UMR5822, Institut de Physique des 2 Infinis de Lyon (IP2I)}
\altaffiliation[Also at ]{Department of Physics, University of Johannesburg, PO Box 524, Auckland Park 2006, South Africa}

\author{Shahram Vatani}
\affiliation{
Galileo Galilei Institute for Theoretical Physics,
Largo Enrico Fermi 2, I-50125 Firenze, Italy
}%
\affiliation{Centre for Cosmology, Particle Physics and Phenomenology, Université catholique de Louvain, Louvain-la-Neuve B-1348, Belgium}


\begin{abstract}
We propose a new family structure for the Standard Model fermions, where the muon is assigned to the third family, taking the placeholder from the tau lepton. This reassignment, which is a mere choice of convention in the Standard Model, becomes physically meaningful in the presence of new physics assuming a direct link between quarks and leptons. In fact, when quark and leptons are coupled by new interactions, the choice of which lepton is assigned to a particular quark generation brings physical consequences, revealing potentially meaningful patterns in the masses and mixings, while pointing to precise and testable predictions for experiments.
\end{abstract}

\maketitle


\section{Introduction}

The Standard Model (SM) of particle physics \cite{Glashow:1961tr,Weinberg:1967tq,Salam:1968rm} has been formulated in the 60's around the principle of gauge invariance \cite{Yang:1954ek}. The electroweak sector of the theory consists of an SU(2)$_L \times$ U(1)$_Y$ gauge symmetry, while an SU(3)$_c$ gauge symmetry is added to account for Quantum ChromoDynamics (QCD) \cite{Fritzsch:1973pi}. One peculiarity is the parity-violating nature of the weak interactions, incarnated in the fact that the SU(2)$_L$ gauge bosons only couple to left-handed leptons \cite{Weinberg:1967tq}.  Quarks respect the same rule as they are also charged under the weak interactions. This forces the fermion content of the SM to be chiral, in the sense that left and right-handed components of the same particles do couple differently to the weak gauge bosons. As a consequence, the lightest states of each type, namely the electron, the electron-neutrino, and the quarks up and down, come together as a ``family'' or ``generation'' of fermions with quantum numbers chosen to cancel gauge and gravity quantum anomalies \cite{Adler:1969gk,Bardeen:1969md,Bell:1969ts,Delbourgo:1972xb,Witten:1982fp}. In fact, a SM family is the minimal anomaly-free chiral set of fields \cite{Geng:1989tcu,Minahan:1989vd,Babu:1989ex}. The breaking of the electroweak symmetry down to Quantum ElectroDynamics (QED) via a Brout-Englert-Higgs (BEH) scalar field \cite{Englert:1964et,Higgs:1964pj} has the benefit of both explaining the Fermi theory of weak interactions and providing mass to all fermions (except for the neutrino).
The latter is obtained via the Yukawa sector of the theory, i.e. the most general set of renormalisable couplings of the scalar field with pairs of fermion fields.

Down to this stage, the construction of the SM and of one family of fermions is only and purely driven by the gauge principle. However, the presence of the muon discovered in 1936 \cite{Muondisc1,Muondisc2} and of the strange quark postulated in 1953 \cite{strange1,strange2}, and later of other heavier leptons and quarks (tau lepton \cite{taulepton}, charm quark \cite{charm1,charm2,charm3}, etc.) revealed the presence of three families, i.e. three copies of the same anomaly-free combination of fields. The measurement of the $Z$ boson width and of the Higgs boson couplings to gluons and photons has ruled out the possible presence of additional chiral families. It is remarkable that the presence of three families is the only aspect of the SM that does not derive from the gauge principle. Via the Yukawa couplings, fermions in different families receive hierarchical masses and non-trivial mixing appears in the $W^\pm$ couplings via the Cabibbo-Kobayashi-Maskawa (CKM) \cite{Cabibbo1963,Kobayashi1973} and Pontecorvo-Maki-Nakagawa-Sakata (PNMS) \cite{Pontecorvo:1967fh,Maki1962} matrices for quarks and leptons, respectively. The current knowledge of the parameters in the Yukawa sector is summarised in Table~\ref{tab:summ} \cite{pdg22}.
The three family structure of the SM, and the patterns in the Yukawa couplings, have been intensively investigated with the purpose of understanding its origins, with notable attempts to restore the gauge principle by gauging family symmetries \cite{Wilczek:1978xi,Ong:1978tq,Maehara:1979kf,Davidson:1979wr,Wu:1979nq,Albrecht:2010xh,Grinstein:2010ve,DAgnolo:2012ulg}, however, no fully satisfactory explanation has been found so far.

\begin{table}[ht]
\caption{\label{tab:summ}%
Summary of the current knowledge of the parameters in the Yukawa sector, expressed in terms of the quark and lepton masses in MeV, and the mixing angles in the CKM and PNMS matrices, expressed in the Chau-Keung parameterisation \cite{ChauKeung1984}. CP-violating phases are not listed, as they are of no relevance for the present work. \\}
\begin{ruledtabular}
\begin{tabular}{lccr}
\textrm{Quarks}&
\textrm{up-type}&
\multicolumn{1}{c}{\textrm{down-type}}&
\textrm{mixing}\\
\colrule
$u$--$d$ & $2.16(\frac{49}{26})$ & $4.67(\frac{48}{17})$ & $s_{\theta_{12}} = 0.22500(67)$\\
$c$--$s$ & $1\ 270(20)$ & $93(\frac{9}{3})$ & $s_{\theta_{23}} = 0.04182(\frac{85}{74})$\\
$t$--$b$ & $172\ 690(300)$ & $4\ 180(\frac{20}{30})$ & $s_{\theta_{13}} = 0.00369(11)$\\
\colrule
\textrm{Leptons} & \textrm{up-type\footnote{For neutrinos, we only know the mass differences from oscillation experiments: $\Delta m_{21}^2 = 7.55(\frac{20}{16}) 10^{-5}$~eV$^2$ and $\Delta m_{32}^2 = 2.42(3) 10^{-3}$~eV$^2$. The neutrino parameters are taken from Ref.~\cite{deSalas:2017kay} for normal ordering, similar values are obtained for inverted ordering and other fits.}}&
\multicolumn{1}{c}{\textrm{down-type}}&
\textrm{mixing}\\
\colrule
$\nu_1$--$e$ & $<10^{-6}$ & $0.511$ & $s_{\theta_{12}} = 0.566(\frac{18}{14})$\\
$\nu_2$--$\mu$ & $<10^{-6}$ & $105$ & $s_{\theta_{23}}=0.740(\frac{14}{20})$\\
$\nu_3$--$\tau$ & $<10^{-6}$ & $1\ 777$ & $s_{\theta_{13}}=0.147(\frac{3}{2})$\\
\end{tabular}
\end{ruledtabular}
\end{table}

At this stage it is important to clarify how families are defined. In fact, at the Lagrangian level, all three copies of the fermionic fields share precisely the same quantum numbers under the SM gauge symmetry, hence they can be freely rotated independently without violating gauge invariance. The SM fermion sector has, therefore, an extended $\text{U}(3)^5$ global symmetry (without counting right-handed neutrinos, whose existence is not confirmed experimentally). A this point, a definition of family is completely arbitrary, subjected to the above global symmetry. However, Yukawa couplings break such symmetry and can be used to define a more physical basis for families. The mass hierarchy among the charged fermions indicates a hierarchical breaking of $\text{U}(3)^5$ to smaller groups and guides a family definition based on the mass eigenstates rather than on generic gauge eigenstates. In the following we adopt this strategy, hence defining families in terms of (approximate) mass eigenstates. We remark that this choice is well justified in models of new physics related to the breaking of the electroweak symmetry. In such a case, the new states prefer to couple to the heaviest fermions of the SM. Hence, the new physics couplings may prefer a specific linear combination of the SM fermion fields, hence providing a second definition of families based on new physics couplings. The latter is expected to be approximately aligned to the hierarchical masses, if the new physics is also related to the breaking of the electroweak symmetry. Following this discussion, the standard definition of families, therefore, groups the SM fermions based on the hierarchy of masses, as illustrated in Table~\ref{tab:summ}. In this work we propose and explore the consequences of an alternative assignment, where muon and tau leptons are exchanged. 

The paper is organised as follows: In section \ref{sec:families} we provide more details on the standard family assignment and the main rationale behind our proposal where the muon is reassigned to the third family, taking the place of the tau. As we will see, this reveals potentially meaningful patterns in the masses and mixing: a new mass pattern emerges pointing to a different origin for the Yukawa couplings of the third generation and the other two. We then propose, in section \ref{sec:masses}, a simple toy model in which the new third generation hierarchy is generated by loops starting from the tree-level Yukawa coupling of the top only. In section \ref{sec:PS} we discuss an explicit example of flavourfull new physics in which the new family structure matters, bringing different results and perspectives when compared to the more standard assignment.
We briefly discuss new experimental consequences at colliders, like the LHC and some future projects, in section~\ref{sec:colliders}, before offering our conclusions in section~\ref{sec:conclusions}.

\section{Families, symmetries and mass hierarchies}
\label{sec:families}

Let us reconsider the rationale behind the standard family assignment in the SM.  As one can see from the current knowledge of the fermion sector 
in Table~\ref{tab:summ}, 
charged fermions show a clear mass hierarchy between particles of the same charge. Furthermore, the mixing angles among quark families are small, signifying that the $W$ boson couples dominantly to well defined couples of quarks: $u$--$d$, $c$--$s$ and $t$--$b$, also following the mass hierarchy. It is therefore natural to associate each pair to a family, with ordering following the masses. This rationale is, however, less evident for leptons, as the mixing angles between families are clearly not hierarchical. Furthermore, in the SM there are no direct connections between quarks and leptons besides the cancellation of anomalies: no interactions allow to clearly associate a family of quarks with a charged lepton. Henceforth, the only guiding principle is the mass, yielding the following family associations: $e$--$u$--$d$, $\mu$--$c$--$s$, $\tau$--$t$--$b$. We want to stress here that this family association within the pure SM realm is completely arbitrary and conventional. While the $e$--$u$--$d$ association makes physical sense as those are the only charged fundamental fermions we observe in ordinary matter, the fermions in the other two families decay down to the first family states. Hence, considering a different assignment, where the muon is associated to the third family instead of the second, is a relevant question to ask.

While within the SM this reassignment is a purely academic exercise, in models of new physics this reassignment can have physical meaning. This is the case if particles interacting with both quarks and leptons are present, as for instance lepto-quarks. They are bosons, of spin zero or one, coupling to one quark and one lepton, hence carrying both baryon and lepton number global charges. Lepto-quarks arise, for instance, in models of Grand Unification \cite{Georgi:1974sy}, and in Pati-Salam models \cite{Pati:1974yy}. In the latter, leptons are associated to quarks by extending the QCD gauge group to SU(4), hence embedding the lepton number by means of a fourth colour. Pati-Salam can be an intermediate step in SO(10) models \cite{Fritzsch:1974nn}. In all these models, quarks and leptons must be embedded into the same representation of the larger gauge groups, therefore  the association within families becomes physically meaningful. In this work we propose the following novel family assignment:
\begin{eqnarray}
1^{\rm st}\; \rm{family}\;\; &\Rightarrow &\;\; e-u-d\,, \nonumber\\
2^{\rm nd}\; \rm{family}\;\; &\Rightarrow &\;\; \tau-c-s\,, \\
3^{\rm rd}\; \rm{family}\;\; & \Rightarrow &\;\; \mu-t-b\,. \nonumber
\end{eqnarray}
We mainly focus on two aspects of this new family structure:
\begin{itemize}
    \item[a)] A new mass pattern emerges, pointing to a different origin for the Yukawa couplings of the third generation and the other two.
    \item[b)] In models with light new physics, sizeable effects are predicted in muon couplings, without the need for lepton flavour violation.
\end{itemize}

\begin{figure}[ht]
\includegraphics[scale=0.85]{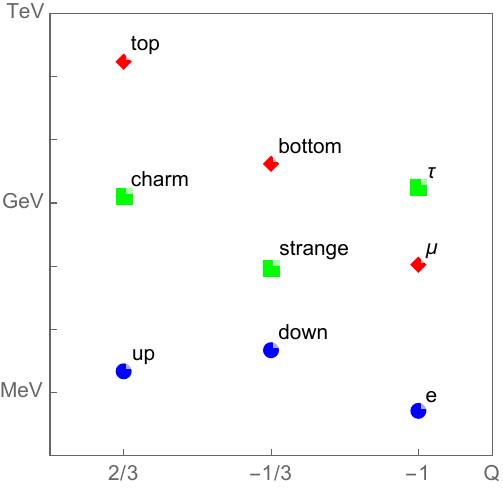}
\caption{\label{fig:masses} Mass ordering for the quarks and leptons in the SM, depending on their electric charge. Colours label the new family assignment we propose.}
\end{figure}

We first discuss the new mass pattern appearing in the new family arrangement, as shown in Fig.~\ref{fig:masses}. First, we observe that the fermions in the first and second families, in blue and green respectively, have similar masses within one order of magnitude, pointing to the MeV scale and GeV scale, respectively. On the other hand, the third family, in red, shows a marked hierarchy between top, bottom, and muon masses. Numerically, the two mass ratios
\begin{align}
 X_{\rm tb}=\frac{m_{\rm t}}{m_{\rm b}} = 41.31^{+0.31}_{-0.21}\,, & \nonumber \\ X_{\rm{b}\mu} = \frac{m_{\rm b}}{m_{\mu}} = 39.56^{+0.28}_{-0.19}\,, &
\label{eq:ratios}
\end{align}
are surprisingly close. This can be better visualised via the ratio of ratios
\begin{equation} \label{eq:ratioofratios}
    \xi = \frac{X_{\rm{b}\mu}}{X_{\rm tb}} = 0.958^{+0.014}_{-0.009}\,,
\end{equation}
which is equal to one within 5\%. This observation suggests that the origin of the Yukawa couplings for the third generation, and for the other two, may be very different. The most likely scenario has the following features:
\begin{itemize}
    \item A family symmetry SU(2) acting on the first two generations. This is broken by effective Yukawa couplings of the order $y_2 \approx 10^{-2}$ and $y_1 \approx 10^{-5}$. 
    \item The third generation has tree-level Yukawa couplings $y_3 \approx 1$, which prevents a larger family symmetry SU(3).
    \item Only the Yukawa coupling of the top quark is allowed at leading order ($y_t \sim 1$), while some new interactions communicate the electroweak symmetry breaking to bottom and muon, explaining their lighter mass.
\end{itemize}
The mass ratio $X \sim X_{\rm tb} \sim X_{\rm{b}\mu}$ is, therefore, generated by the latter mechanism. 

\section{A simple toy model for the masses}
\label{sec:masses}
We envision two model-building avenues: the Froggatt-Nielsen mechanism \cite{Froggatt:1978nt}, where lighter fermions inherit the top Yukawa via mixing to heavy fermions, and loop mass generation. The latter has been first used in attempts to explain the electron mass from the muon one in the early 70's \cite{Georgi:1972hy}. Attempts to generate all quark and lepton masses from the top Yukawa date back to the 90s \cite{He:1989er,Babu:1990vx,Rattazzi:1990wu}, but mainly failed to generate the lightest masses and to pass flavour changing neutral current (FCNC) tests. 
Our scenario is free from such issues, as only the third generation masses would need to be loop generated.

A bottom-up approach reveals that there is a unique minimal model with two additional scalars that can generate the bottom and muon masses out of one and two loops, respectively, from the top Yukawa. The proposed scalars are triplets under QCD, and carry the following weak isospin and hypercharge, ${\bf I}_Y$: $S = {\bf 1}_{-1/3}$ and $\phi = {\bf 2}_{1/6}$. Interactions with the SM fermions are generated at a high scale $\Lambda$, in the form of the most general Yukawa-like couplings allowed by the gauge principle:
\begin{align}
    \mathcal{L}_{\rm Yuk} = & y_t\ \bar{t}_R q_L\ H + \lambda_S\ S\  \Big( c_{qq}\ \bar{q}^c_L q_L + c_{tb}\ \bar{b}^c_R t_R \nonumber \\
    & + c_{ql}\ \bar{q}_L l_L^c + c_{t\mu}\ \bar{t}_R \mu_R^c  + c_{b\nu}\ \bar{b}_R \nu_R^c\Big) + \nonumber\\
    & \lambda_\phi\ \phi\ \Big( c_{q\nu}\ \bar{q}_L \nu_R + c_{bl}\ \bar{b}_R l_L \Big) + \mbox{h.c.}  
\label{Yukawa_loops}
\end{align}
where the coefficients $c_{ij}$ depend on the ultra-violet completion of the model, and $H$ is the BEH doublet. We have included right-handed neutrinos $\nu_R^i$ with a Majorana mass, which will be needed to implement the see-saw mechanism to suppress neutrino masses. 
The only model-dependent assumptions, which should be justified at the level of the ultra-violet completion, are: only the top Yukawa coupling is present, while $y_{b,\mu}(\Lambda) = 0$; only the coefficient $c_{ql} (\Lambda) = 0$, while a non-zero value is generated by loops of $\phi$ to yield the correct suppression in the muon mass. 
Defining
\begin{equation}
    \epsilon_{S/\phi} = \frac{1}{8 \pi^2} \ln \frac{\Lambda}{M_{S/\phi}}\,,
\end{equation}
at the scalar mass scale
\begin{equation}
    c_{ql} = \lambda_\phi^2 \ n_R \ \epsilon_\phi \cdot (c_{q\nu} c_{b\nu} c_{bl}^\ast)\,,
\end{equation}
where $n_R$ is the number of right-handed neutrinos lighter than the scalars.
Both bottom and muon masses are generated at one loop from the top mass, however $m_\mu$ contains an extra suppression via $c_{ql}$. Including only the logarithmic contribution, corresponding to the renormalisation group evolution from $\Lambda$ down to the scalar mass scales, we find:
\begin{align}
   &  \frac{m_{\rm b}}{m_{\rm  t}} = N_c\ \lambda_S^2\ \epsilon_S \cdot (c_{tb}^\ast c_{qq})\,, \nonumber \\ & \frac{m_\mu}{m_{\rm b}} = n_R\ \lambda_\phi^2\ \epsilon_\phi \cdot \frac{c_{t\mu}^\ast c_{q\nu}^\ast c_{b\nu}^\ast c_{bl}}{c_{tb}^\ast c_{qq}}\,, 
\end{align}
where $N_c = 3$ is the number of QCD colours. These values should match the mass ratios at a scale corresponding to the scalar masses. A Dirac neutrino mass is also generated from the bottom mass via a loop of $\phi$, giving
\begin{equation}
    \frac{m_\nu^D}{m_{\rm b}} = N_c\ \lambda_\phi^2\ \epsilon_\phi \cdot (c_{q\nu^\ast} c_{bl})\,.
\end{equation}
Hence, the Majorana mass for the light ``third-family'' neutrino is given by the see-saw formula:
\begin{equation}
    m_\nu = \frac{(m_\nu^D)^2}{M_R} = N_c^2 \lambda_\phi^4 \epsilon_\phi^2 \frac{m_{\rm b}^2}{M_R} \cdot (c_{q\nu^\ast} c_{bl})^2\,,
\end{equation}
where $M_R$ is the Majorana mass of the right-handed neutrinos.
These estimates are valid for the mass hierarchy $M_R \lesssim M_\phi \sim M_S$. By matching these estimates with the values of the ratios $X_{\rm tb}$ and $X_{\rm{b}\mu}$ at the scale $M_R$ and the neutrino mass, we can estimate the values of the necessary mass scales. For $m_\nu = 1$~eV,  $n_R = 3$ and $\lambda_S = \lambda_\phi = 0.3$, setting all coefficients $c=1$ we obtain, for instance, 
\begin{equation}
    M_R = 1.3\ 10^7~\mbox{GeV}\,, \quad \frac{M_S}{M_\phi} = 1150\,, \quad \frac{\Lambda}{M_S} = 490\,,
\end{equation}
where the value of the scalar masses cannot be determined, but are higher than the right-handed neutrino masses. Note that the scale $\Lambda$ is large enough to avoid dangerous FCNCs, however the couplings of $S$ violate baryon and lepton number and will generate proton decay via the process $p \to K_0 \mu^+$. The effective scale suppressing this process is required to be $\Lambda_p > 10^{16}$~GeV. Assuming a CKM-like suppression for the mixing of the up and down-type quarks with the third generation, we estimate $M_S/\lambda_S \gtrsim 10^{12}$~GeV. The bound from proton decay could be removed altogether by introducing two $S$ scalars: $S_{qq}$ that couples only to di-quarks and carries baryon number $B = -2/3$, and $S_{lq}$ that couples to one lepton and one quark and has $B=1/3$ and $L=1$. Note, finally, that the large scale is mainly due to the neutrino masses and the seesaw mechanism. It is possible to lower the right-handed neutrino masses $M_R$ by employing the inverse seesaw mechanism \cite{Mohapatra:1986aw}, hence lowering the scalar masses closer to the TeV scale.

In this class of models, the flavour mixing patterns will crucially depend on the mechanism generating the Yukawas for the second and third generation. An attractive possibility would be to use loops to communicate the third generation Yukawas to the other two, however this programme has met crucial difficulties \cite{Weinberg:2020zba}. A detailed analysis of this question is left for future investigations.

\section{Flavour in a Pati-Salam example}
\label{sec:PS}
The second consequence of assigning the muon to the third family could emerge in the low energy effects of light new physics, around the TeV scale. This is justified by the inherent link between the electroweak symmetry breaking and the top quark, who enjoys the largest coupling to the BEH field in the SM. Henceforth, new physics at the TeV scale may couple preferentially to third generation fermions. New physics coupling dominantly to the muon has been hinted at by recent measurements (see the recent review \cite{DAlise:2022ypp}), mainly stemming from $B$ meson decays and the measurement of the muon anomalous magnetic moment. 
If the new states were to couple dominantly to third generation, following the usual assignment, a sizeable mixing between tau and muon would be required to generate large effects involving muons, hence leading to lepton-flavour violation (LFV) \cite{Calibbi:2015kma,Guadagnoli:2022oxk}. LFV, however, is strongly constrained by experiments at levels well beyond the constraints on lepton-flavour universality violation (LFUV). If the muon is assigned to third generation, anomalies involving muons could be explained without the need for LFV signatures \cite{Alonso:2015sja}, hence avoiding the most stringent constraints on the models.
As a minimal model, we will consider here a single spin-1 lepto-quark, $U_1 = {\bf 1}_{2/3}$, which could emerge from Pati-Salam (PS) unification \cite{DiLuzio:2017vat}. If the muon belongs to the same multiplet as the bottom, like in our model assumption, the couplings
\begin{equation}
    \mathcal{L}_U = U_1^\mu \  \bar{b} \left(\kappa_L P_L + \kappa_R P_R\right) \mu + \mbox{h.c.}
\end{equation}
can be generated directly, without the need of a sizable mixing between muons and taus. Hence, LFV can be absent or naturally strongly suppressed in this class of models. This kind of new physics was mainly motivated by an early observation of LFUV in B-meson decays \cite{LHCb:2021trn}, which however was shown to be due to systematics in a follow-up reanalysis of the data by the same collaboration \cite{LHCb:2022qnv,LHCb:2022zom}. Nevertheless, lepto-quarks like $U_1$ emerge in motivated models and could generate in the future signals of new physics, which are worthy searching for at collider experiments. 

As a concrete higher energy realisation, in the following we will consider a variation of the PS$^{3}$ model of Refs~\cite{Bordone:2017bld,Bordone:2018nbg,Cornella:2021sby}. The model is based on family-specific PS symmetries, broken at different energy scales. As a consequence, the lightest $U_1$ lepto-quark couples dominantly with one of the PS-families. In the original proposal, the masses of the PS-families are also hierarchical, hence the lepto-quark naturally couples to the heaviest fermions: top, bottom and tau.
To explain the emergence of sizeable effects on the muon, an additional coupling of the lepto-quark to the second PS-family was included, hence introducing a new parameter in the model.

Following the principle of assigning the muon to the third family, we propose an alternative version of the PS3 model, where the third PS-family does not receive the largest mass. To better understand the difference, without entering too much into the details of the model that are explained in the original articles, we recall the generic mass matrix for the SM fermions, which has the following general structure:
\begin{equation}
    \mathcal{L}_{\rm mass} = - \bar{f}_R \left( \begin{array}{ccc} \epsilon_{LR} U_{11} & \epsilon_{LR} U_{12} & 0 \\
    \epsilon_{LR} U_{21} & \epsilon_{LR} U_{22} & \epsilon_L \\
    0 & \epsilon_R & y_f v \end{array} \right)f_L\,,
\end{equation}
\begin{figure}[t!]
    \centering
    \includegraphics[scale=0.55]{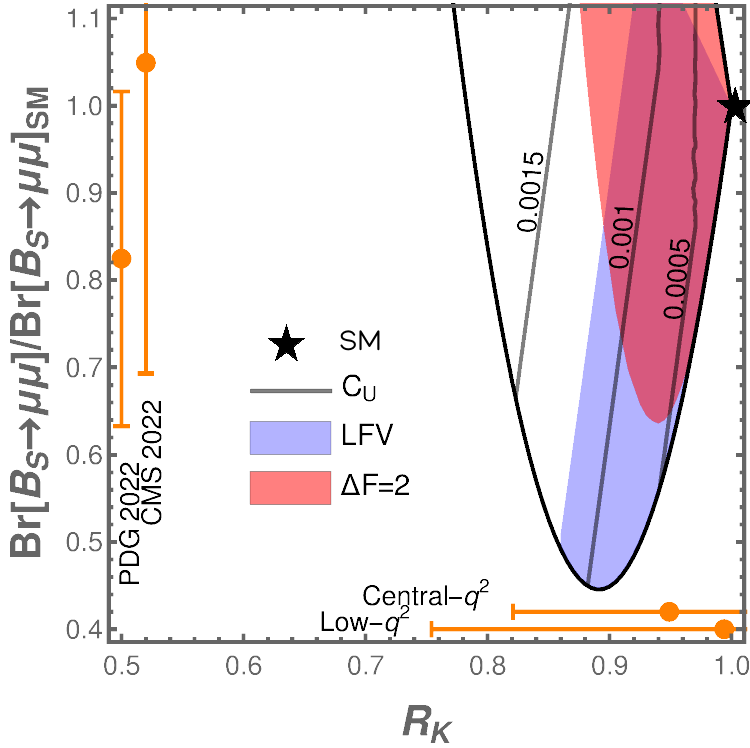}
    \caption{ \label{fig:PS3} Allowed parameter space in the PS$^{3}$ model with muon in the third generation. The parameter space is delimited by the solid black line, with constrains from LFV ($\tau \to 3 \mu$) in blue \cite{Hayasaka:2010np,HFLAV:2022pwe} and $\Delta F = 2$ bounds in red \cite{HFLAV:2022pwe}. Contours for the lepto-quark coupling $C_U$ are indicated in grey and decrease to the decoupling limit, reaching the SM prediction represented by the black star. The orange points along the axes show the most recent experimental measurements for $R_K$ \cite{LHCb:2022zom} and $B_S \to \mu^+ \mu^-$, where we show both the most recent CMS result \cite{CMS:2022dbz} and the previous World average \cite{pdg22}.}
\end{figure}
where the $\epsilon$ terms, linear in the Higgs vacuum expectation value, are generated by higher dimensional operators, stemming from the breaking of the first and second generation PS symmetries, while $y_f$ comes from a renormalisable coupling of the BEH field to the third PS-family. Hence, a natural  hierarchy $\epsilon_R < \epsilon_{LR} < \epsilon_L < y_f$ emerges from the model. The $2\times 2$ matrix $U$ parameterises the breaking of the U(2) symmetry between the first two PS-families. In this model, the new assignment of the muon to the third generation is equivalent to assuming that $\epsilon_L \sim m_\tau$, while $y_f v \sim m_\mu$. Note that $\epsilon_L \sim$~GeV is consistent with the masses of the second generation quarks, which stem from the second PS-family. However, the assumption $y_f v \ll \epsilon_L$ implies that:
\begin{itemize}
    \item[a)] The left-handed muon belongs to the third PS-family; \item[b)]  The right-handed muon belongs to the second PS-family.
\end{itemize}
The PS lepto-quark $U_1$ that couples to the third PS-family, therefore, will dominantly couple to the left-handed muons and the right-handed taus. 

We re-analysed the current low energy limits in the new configuration, which naturally predicts sizeable effects on the muon couplings. We find that the bounds on the model parameter are very similar to the original model, except for observables involving only muons. 
We focus here on a benchmark in the parameter space characterised by the $b\to s \mu^+ \mu^-$
processes, incarnated in the ratios 
 \begin{equation}
    R_{K^{(*)}} = \frac{\mbox{Br} [B \to K^{(*)} \mu^+ \mu^-]}{\mbox{Br} [B \to K^{(*)} e^+ e^-]}\,,
 \end{equation}
and $\mbox{Br} [B_S \to \mu^+ \mu^-]$ \cite{pdg22} normalised to the SM central value
\begin{equation}
    \left. \mbox{Br} [B_S \to \mu^+ \mu^-]  \right|_{\rm SM} = (3.65 \pm 0.23) \cdot 10^{-9}\,.  
\end{equation}
We recomputed the bounds on the PS$^3$ model under the assumption of third generation muon, i.e. $\epsilon_L \sim m_\tau$, showing the allowed region for a specific benchmark model in Fig.~\ref{fig:PS3}. The black line contours indicate the predicted region, while the red and blue areas indicate the regions preferred by $\Delta F = 2$ constraints (mainly $B_S -\bar{B}_S$ mixing)  \cite{HFLAV:2022pwe} and LFV ones (mainly $\tau \rightarrow 3 \mu$) \cite{Hayasaka:2010np,HFLAV:2022pwe}, respectively, while the bounds from LFUV do not constrain the allowed region. 
The intersection of the two regions indicate the prediction of the model to be compared to the experimental determinations of $R_K$ \cite{LHCb:2022qnv,LHCb:2022zom} and $\mbox{Br} [B_S \to \mu^+ \mu^-]$ \cite{CMS:2022dbz,pdg22}, indicated by the orange points in the plot (the error bands take into account the experimental errors only).
Compared to the original model, in our set up the new physics effects are driven by the coupling $g_U$ of the lepto-quark, which is expressed in terms of 
\begin{equation}
    C_U = \frac{g_U^2 v^2}{M_U^2} \,, 
\end{equation}
where $v=246$~GeV is the SM vacuum expectation value, $M_U$ and $g_U$ are the lepto-quark mass and coupling, respectively. For values of $g_U \sim 3$ as used in \cite{Bordone:2017bld}, this leads to $M_U \sim 10$~TeV, however this coupling is only required to be larger than 1 hence $M_U \sim$ few TeV is possible. The presence of a lepto-quark $U_1$ from the PS3 model will be tested both at the LHC and at future high-energy colliders.

\section{Third-family muons at colliders} \label{sec:colliders}

Under the hypothesis that new physics couples dominantly to third generation fermions, the new proposed family structure would predict naturally muon-rich final states at colliders.

Lepto-quarks are a neat example, as they will decay into top-bottom and muons, a final state that would have been considered inter-generation. However, searches have been implemented, mainly motivated by the LFUV anomalies in the B-meson sector. As an example, we consider here the vector lepto-quark $U_1$ from the previous section. The minimal model with left-handed couplings (LH $U_1$) will predict a 50\% rate into $t\nu$ and a 50\% in $b$ plus a charged lepton. The latter will be the tau for the standard family assignment, or the muon in the new proposal. Both configurations haven been searched for by ATLAS and CMS collaborations at the LHC. The main production mode is due to pair production via QCD interactions, which leads to a cross section that only depends on the mass of the lepto-quark. For a fair comparison, we consider here only searches employing the largest dataset, corresponding to $\sim 140$~fb$^{-1}$ at $13$~TeV in the centre of mass. For the tau option, ATLAS ~\cite{ATLAS:2022wcu} was able to exclude masses above $1740$~GeV, while CMS~\cite{CMS:2022dbz} excludes masses above $1700$~GeV for a generic search tagging taus. A more recent note unveiled a dedicated search for the $U_1$ lepto-quark, which also includes other production modes \cite{CMS-PAS-EXO-19-016}. In the latter, pair production leads to an exclusion above $1860$~GeV: interestingly, however, the search reveals an excess in the data at high invariant mass, which is quantified to circa $3\sigma$ for a mass hypothesis of $2$~TeV. Furthermore, for couplings larger than one, t-channel contribution to the process $b\bar{b}\to \tau^+ \tau^-$ become relevant, raising the mass limit up to $2500$~GeV for a coupling equal to $2.5$. The muon channel have been studied by ATLAS \cite{ATLAS:2023uox}, leading to a stronger bound of $1980$~GeV. A summary of these results is  presented in Table~\ref{tab:LQatcolliders}. Clearly muons are easier to tag at the LHC, however final states with taus also lead to competitive bounds.

\begin{table}[t!]
\caption{\label{tab:LQatcolliders}%
Current LHC limits for the $U_1$ vector lepto-quark in various models: an effective model with left-handed couplings (LH $U_1$) and the PS3 model. We compare the traditional option with the one stemming from the muon in third family.\\}
\begin{ruledtabular}
\begin{tabular}{ccccc}
 & \multicolumn{2}{c}{Tau in 3$^\text{rd}$} & \multicolumn{2}{c}{Muon in 3$^\text{rd}$} \\
\colrule
 & $t\nu:\; 50\%$ & & $t\nu:\; 50\%$ & \\
LH $U_1$ & $b\tau:\; 50\%$ & $1860$~GeV & $b\tau:\; 0\%$ & $1980$~GeV \\
 & $b\mu:\; 0\%$ & & $b\mu:\; 50\%$ &\\
\colrule
 & $t\nu:\; 50\%$ & & $t\nu:\; 50\%$ & \\
PS3 & $b\tau:\; 50\%$ & $1860$~GeV & $b\tau:\; 25\%$ & $\sim 1900$~GeV\\
 & $b\mu:\; 0\%$ & & $b\mu:\; 25\%$ & \\
\end{tabular}
\end{ruledtabular}
\end{table}

We also analysed the case of the PS3 lepto-quark, which can couple both to left- and right-handed fermions. As the right-handed neutrino is light in this model \cite{Bordone:2018nbg}, the final state decays are equally represented by $t\nu$ and $b\tau$ in the original proposal. In our alternative model, the left-handed coupling only involved muons instead of taus, hence leading to branching ratios of 50\%--25\%--25\% in $t\nu$--$b\tau$--$b\mu$, respectively, as summarised in Table~\ref{tab:LQatcolliders}. The current limit is hard to estimate for the latter case, however we expect that it will be mainly driven by the muon final state, and be close to $1900$~GeV \cite{ATLAS:2023uox}

Future hadron collider projects at higher energy will be able to probe much higher mass scales, reaching above $10$~TeV for the $100$~TeV option \cite{Allanach:2019zfr}. However, the muon assignment to the third generation is particularly interesting for the recent proposal of a high energy muon collider, which would then be an effective third-generation collider and could lead to a direct probe of new physics that couples to third generation fermions. The case of lepto-quarks coupling to muons has been recently studied \cite{Asadi:2021gah}, showing that interference with Drell-Yan production of two b-quarks can probe masses in the multi-tens of TeV for sizeable couplings.




\section{Conclusions}
\label{sec:conclusions}
Generation assignments of fermions in the pure SM is arbitrary, and it may be only driven by the hierarchical mass structure emerging from the Yukawa couplings. However, this is not the case in extensions of the SM where a link in terms of interactions is present among quark and leptons, and flavour is not universal among the fermion generations. 
We have discussed the potential benefits of assigning the muon to the third generation, in the place of the tau lepton, in these extensions of the SM. This leads to physical consequences, which could be tested at experiments. To start, a new understanding of the mass patterns emerges, pointing towards a different mechanism behind the generation of masses in the third generation versus the other two. Furthermore, sizeable deviations in the couplings of muons to the SM can be generated, without the need for lepton flavour violation signatures, which are tightly constrained. Such effects, mainly due to light new physics around the TeV scale, point towards anomalies in low energy measurements involving muons, which will be tested in the near future via high precision measurements of the muon anomalous magnetic moment and rare decays of mesons.

\bibliography{apssamp}

\end{document}